\documentclass[5p,authoryear]{elsarticle}
\makeatletter 
\def\ps@pprintTitle{%
 \let\@oddhead\@empty
 \let\@evenhead\@empty
 \let\@evenfoot\@oddfoot} 
\makeatother
\usepackage[utf8]{inputenc} 
\usepackage[T1]{fontenc}
\usepackage[french]{babel}
\usepackage[babel=true]{csquotes} 
\usepackage[fleqn]{amsmath} 
\usepackage{amsthm} 
\usepackage{booktabs} 
\usepackage{multirow} 
\usepackage{amssymb} 
\usepackage{float}
\usepackage{subfigure}
\usepackage{bm}
\usepackage{adjustbox}
\usepackage{hyphenat}
\usepackage{hyperref} 
\usepackage[french]{cleveref} 

\begin{document}
\begin{sloppypar}

\begin{frontmatter}

\title{Vertebrae localization, segmentation and identification using a graph optimization and an anatomic consistency cycle}

\author{Di Meng\corref{cor1}}
\ead{di.meng@inria.fr}
\author{Edmond Boyer, Sergi Pujades}
\address{Inria, Univ. Grenoble Alpes, CNRS, Grenoble INP, LJK, France}
\cortext[cor1]{Corresponding author:}

\begin{abstract}
Vertebrae localization, segmentation and identification in CT images is key to numerous clinical applications.
While deep learning strategies have brought to this field significant improvements over recent years, transitional and pathological vertebrae are still plaguing most existing approaches as a consequence of their poor representation in training datasets.
Alternatively, proposed non-learning based methods take benefit of prior knowledge to handle such particular cases.
In this work we propose to combine both strategies. To this purpose we introduce an iterative cycle in which individual vertebrae are recursively localized, segmented and identified using deep-networks, while anatomic consistency is enforced using statistical priors. In this strategy, the transitional vertebrae identification is handled by encoding their configurations in a graphical model that aggregates local deep-network predictions into an anatomically consistent final result. 
Our approach achieves state-of-the-art results on the VerSe20 challenge benchmark, and outperforms all methods on transitional vertebrae as well as the generalization to the VerSe19 challenge benchmark.
Furthermore, our method can detect and report inconsistent spine regions that do not satisfy the anatomic consistency priors.
Our code and model are openly available for research purposes\footnote{}.
\end{abstract}

\begin{keyword}
Spine \sep Vertebrae \sep Segmentation \sep Labelling \sep Transitional vertebrae\end{keyword}

\end{frontmatter}

\footnotetext{\url{https://gitlab.inria.fr/spine/vertebrae_segmentation}}


\newcommand{\todo}[1]{{\bf{\color{red} TODO: #1}}}

\newcommand{\mat}[1]{\vect{#1}}
\newcommand{\matelem}[2]{\mat{#1}_{#2}}
\newcommand{\vecelem}[2]{\vect{#1}_{#2}}
\newcommand{\argminD}{\arg\!\min}

\newcommand{\eqnref}[1]{Eq.~(\ref{#1})}
\newcommand{\sectref}[1]{Sec.~\ref{#1}}
\newcommand{\figref}[1]{Fig.~\ref{#1}}
\newcommand{\tabref}[1]{Tab.~\ref{#1}}
\newcommand{\eg}{e.g.~}
\newcommand{\ie}{i.e.~}

\newcommand{\R}{\mathbb{R}}

\newcommand{\smpl}{Sp}                   
\newcommand{\smplsurf}{\mathcal{Sp}}                   

\newcommand{\bweights}{\mat{W}}

\newcommand{\trans}{\vect{t}}                     
\newcommand{\rot}{\vect{r}}                     
\newcommand{\ff}{\vect{f}}                     
\newcommand{\pose}{\boldsymbol{\theta}}                 
\newcommand{\shape}{\boldsymbol{\beta}}                 

\newcommand{\vol}{\mat{V}} 
\newcommand{\scan}{\mathcal{S}} 

\newcommand{\vt}{\mat{T}}                        
\newcommand{\algninit}{\mathcal{I}}                      
\newcommand{\algnshape}{\mathcal{A}}                      

\newcommand{\vtmean}{\vt^\mathrm{\mu}}                        
\newcommand{\shapespace}{\mat{B}}                      

\newcommand{\vtGT}{\mat{T}_\mathrm{GT}}                 
\newcommand{\vtEst}{\mat{T}_\mathrm{Est}}			   
\newcommand{\vtCloth}{\mat{T}_\mathrm{cloth}}			   
\newcommand{\vtFusion}{\mat{T}_\mathrm{Fu}}

\newcommand{\fusionscan}{\mathcal{S_\mathrm{Fu}}}
\newcommand{\scanvertex}{\vect{s}} 

\newcommand{\joints}{\mat{J}}
\newcommand{\posebs}{B_p(\pose)}          
\newcommand{\shapebs}{B_s(\shape)}          
\newcommand{\posebsvertex}{b_p(\pose)}          
\newcommand{\shapebsvertex}{b_s(\shape)}          
\newcommand{\vtvertex}{\vect{v}}   

\section{Introduction}
\label{intro}
 
Segmenting, identifying and localizing vertebrae in CT images play an essential role in many clinical applications such as the spinal deformities assessment \citep{forsberg2013fully} or computer-assisted surgical interventions \citep{knez2016computer,merloz1998computer}.
These tasks are intrinsically interdependent and a common issue comes from inconsistencies that can propagate between them. For instance neighboring vertebrae share similar shapes which makes their identification uncertain. Moreover pathological spines can present abnormal shapes or the number of transitional vertebrae can be different among patients. These transitional vertebrae, \ie the absence of T12 or occurrence of T13 or L6,  are common and are reported to affect between $15\%$ and $35\%$ of the general population~\citep{carrino2011effect,uccar2013retrospective,konin2010lumbosacral}. However, they only impact one vertebra among the $24$ in a spine and their occurrence percentages in standard vertebrae datasets appear to be therefore much lower, down to a few percentages in practice.
As a consequence, the current state-of-the-art methods, which are mostly based on deep networks trained over standard datasets, tend to experience important performance drops in the presence of transitional vertebrae.

Besides augmenting datasets, an alternative solution is to exploit prior knowledge on the full structure of the spine, as initially introduced in non-learning based methods. In this work, we investigate the combination of such a strategy with a deep learning approach. Precisely we propose to iteratively cycle between the localization, segmentation and identification tasks, which uses deep networks, while enforcing anatomical consistency with statistical priors. 
On one hand, the priors are used to localize vertebrae: they leverage learned statistics of vertebrae volume and inter-vertebral distances which exhibit more robustness to pathological cases than shape or appearance models.
On the other hand, in order to handle transitional vertebrae in the identification, we propose to encode the admissible configurations in a graphical model. Such a model leverages local deep-network predictions and aggregates them into an anatomically consistent result. Our experiments demonstrate that this strategy successfully handles local inaccurate predictions. Therefore it performs better than other methods with transitional vertebrae while providing state-of-the-art results on standard benchmarks such as VerSe20. 
The contributions of this paper are summarized as follows:
\begin{itemize}
\item We propose an anatomic consistency cycle for the unified task of vertebrae localization, segmentation and identification. The three tasks cooperate and complement each other, achieving an anatomically coherent result. In case the anatomic consistency criteria were not met, the region of inconsistency is reported.
\item We propose to leverage the statistical shape priors with deep neural networks. The rare cases, such as transitional vertebrae and pathologies which are hard to be learned with the limited representation in the training set, are well detected.
\item We extend the work on individual vertebra classification~\citep{mohammed2020morphology} and show significant improvement by including neighboring information. We experimentally quantify that solely using the shape of the vertebrae achieves higher or comparable performance than using as input the CT image.
\item We introduce a graph to enforce the global consistency of the predicted individual vertebrae labels. It allows to detect anatomically rare cases, e.g.\ the transitional vertebrae T13, L6 and the absence of T12.
\end{itemize}

\section{Related work}
We start by reviewing the vertebrae labelling literature, addressing the localization and identification task.
Then we review the multi-label segmentation methods which in addition includes the vertebrae segmentation.

\subsection{Vertebrae labelling}
For vertebrae localization and identification, traditional methods build on shape models \citep{cai2015multi,lindner2014robust} or graphical models \citep{glocker2012automatic,bromiley2016fully} to detect landmarks.
In \citep{glocker2012automatic} vertebrae locations are regressed by making assumptions on the shape appearance.
Such models are powerful since they encode prior knowledge about the full structure of the spine.
However, surface~\citep{klinder2009automated,aslan20103d,cai2015multi,meng2020learning} or image based appearance models \citep{glocker2012automatic,bromiley2016fully}, which make assumptions on exact shapes, are not robust to pathological cases (such as fractures or metal implants).
Relative positions of vertebrae~\citep{glocker2012automatic,glocker2013vertebrae,wang2021automatic} are shown to be less sensitive to these cases. 
In our work we propose to learn statistics on vertebrae relative positions and volumes.
The volume quantifies a {\em part of the vertebra shape} that is robust to morphological deviations. Furthermore we learn statistics conditioned on the spine level (cervical, thoracic, lumbar), increasing hence their performance. Moreover we relax the need for a fixed threshold to detect vertebrae~\citep{payer2020coarse} and compute adaptive thresholds that automatically vary across spine levels and patients.

More recent methods build on machine learning and convolutional neural networks (CNN) in which hand-crafted features tend to be replaced by learned ones \citep{glocker2013vertebrae,chen2015automatic}. \cite{sekuboyina2018btrfly} combine information across several 2D projections using a butterfly-like architecture and encode the local spine structure as an anatomic prior with an energy-based adversarial training.  \cite{liao2018joint} and \cite{qin2020residual} develop a multi-label classification and localization network using FCN and residual blocks. They improve the classification branch with bidirectional recurrent neural network (Bi-RNN) to encode short and long range spatial and contextual information. In \citep{wang2021automatic} a nnUnet keypoint detection model is trained to estimate $26$ vertebrae activation maps (including the sacrum). \cite{chen2019vertebrae} propose a variant FCN that localizes vertebrae in original resolution and classifies them in down-sampled resolutions. These are stand-alone networks \citep{tao2022spine} that directly output vertebrae locations and identifications and can be trained in an end-to-end manner.
All these methods have difficulties with transitional vertebrae.
The scarcity of data with rare cases makes the end-to-end learning approaches under-perform in their detection.

Other works in the literature perform vertebrae labelling in multiple stages.
\cite{mccouat2019vertebrae} employ separate CNNs to localize the vertebrae in 3D samples and identify them in 2D slices with a two-stage method. In order to find vertebrae, \cite{jakubicek2020learning} detect first the spine and then track the spinal cord based on the combination of a CNN and a growing sphere method. \cite{payer2020coarse} follow a similar strategy of coarse to fine vertebrae localization by detecting the spine in lower resolutions and each vertebra in higher resolution images. For identifying  vertebrae, \cite{chen2015automatic} propose a joint learning model (J-CNN) that can classify the vertebrae labels and encode the pairwise conditional dependency at the same time.
In our work, we take a two-step approach to localize and identify vertebrae. To identify individual vertebra we build on \cite{mohammed2020morphology} and use the segmented vertebrae shape instead of the raw CT image. 

Since raw predictions from networks are not always accurate, various post-processing stages have been adopted to constrain vertebrae locations based on the anatomical ordering.
\cite{chen2015automatic} use a shape regression model to correct the offsets from the deviation in the vertical axis. The model assumes that the coordinates distribution can be described by a quadratic form and is limited to coordinates on the vertical axis. \cite{yang2017automatic} introduce a chain-structure graphical model to depict the spatial relationship between vertebrae and regularize their locations with an $L_1$ norm  to learn the best sparse representation. \cite{chen2019vertebrae} model the score maps interpolated along the 1-D spinal curve with a Hidden Markov Model to generate the optimized 1-D coordinates.
\cite{mader2018detection,mader2019automatically} learn an optimal conditional random field as a spatial regularizer to select a global optimal configuration over all the landmarks.
In our approach, we contribute a new graphical model to select a global optimal configuration among all the individual predictions, within which the transitional vertebrae are explicitly modeled.

\subsection{Multi-label vertebrae segmentation}

For the  segmentation task, earlier methods mainly used statistical models \citep{klinder2009automated, rasoulian2013lumbar}, active shape models \citep{graham1995active,benjelloun2011framework,al2016improving}, graph cuts \citep{aslan20103d}, and level sets \citep{tsai2003shape,lim2014robust}. 
Recent learning based approaches have nevertheless demonstrated better performances. 
\cite{al2017shape} use a deep convolutional neural network for cervical vertebrae segmentation and introduce a shape-aware term in the loss function. The approach is  semi-automatic and the vertebrae locations are given manually.
\cite{sekuboyina2017localisation} propose a two-stage approach for lumbar vertebrae segmentation in which they first regress the bounding box of the lumbar region and then segment and label each vertebra locally. 
In a similar fashion, we compute a binary segmentation of the full spine that is used as strong support in locating the region of interest.

Several methods achieve very good performances by assuming that a specific part of the spine is observed~\citep{janssens2018fully,al2018fully,al2018spnet}. However, these methods are not robust to arbitrary field of views.
\cite{lessmann2019iterative} handle this issue by iteratively applying convolutional networks in the CT images. The vertebrae are segmented and identified as they are progressively found with a sliding window.
This method relies on the first detection and hardly recovers from failures that occur during this first stage. To address this limitation, \cite{altini2021segmentation} propose a semi-automatic method that requires the user to provide the identity of the first top vertebra and the total number of vertebrae.

\cite{masuzawa2020automatic} propose a multi-stage framework where, first,  the bounding boxes of cervical, thoracic and lumbar vertebrae are found, then the vertebrae in each bounding box are segmented and identified in an iterative manner. This approach reduces the intra-class variance from identifying one out of 24 vertebrae into identifying one out of 7 cervical, 12 thoracic and 5 lumbar vertebrae. Similarly, \cite{mohammed2020morphology} quantify that a two-step approach has a higher identification rate. In our work we leverage this two-step approach and extend the work of \cite{mohammed2020morphology} by including neighbouring vertebrae as input. 

\cite{payer2020coarse} use a three-step approach to first localize the spine, then simultaneously locate and identify each vertebra and finally segment each individual vertebra. A fixed threshold is used for missing vertebrae. \cite{kim2021automatic} propose a very similar approach to detect and segment vertebrae in X-ray images. 
In our method we propose an anatomic consistency cycle enforcing a coherent final result. The segmentation module builds on the segmentation network of \cite{payer2020coarse} and the refinement strategy of \cite{lessmann2019iterative}.

\begin{figure*}[t]
    \centering
    \includegraphics[width=0.9\textwidth]{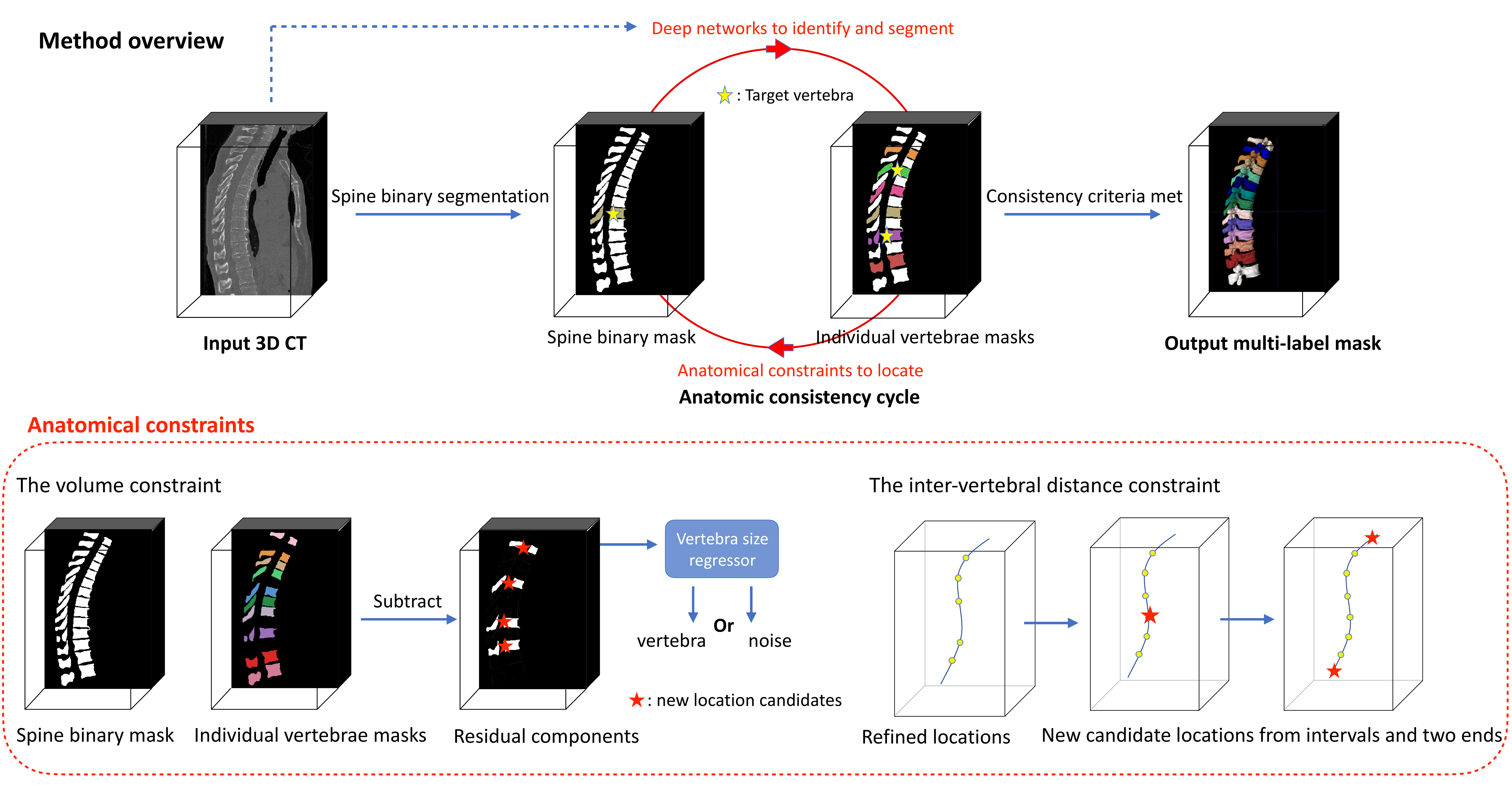}
    \caption{The method overview. Given a 3D CT as input, the spine is segmented as a reference to locate the individual vertebrae. The anatomical constraints are leveraged with deep networks for localization and segmentation. Once the location is stable, its identification is obtained given its segmentation. The set of location, segmentation and identification is cycled through until the consistency criteria are met.}
    \label{fig:method_overview}
\end{figure*}

\section{Method}

We propose to cycle through the three tasks of localization, segmentation and identification in order to enforce anatomic consistencies among them (Sec.~\ref{sec:cycle}), namely consistencies among locations and segmentations (Sec.~\ref{sec:anatomic_constraints}) and their identifications~(Sec.~\ref{sec:vert_identification}).
The overview of the method is shown in Fig~\ref{fig:method_overview}.

The input to our method is a CT volume with an arbitrary field of view, resolution and anatomic position, potentially imaging a human spine. The CT volume is re-sampled into an isotropic resolution of $1\times1\times1$mm\textsuperscript{3} and re-oriented into the same anatomic orientation as pre-processing. The obtained result will be transformed back to the original orientation and resolution.

\subsection{Anatomic Consistency Cycle}
\label{sec:cycle}
Our method starts by segmenting a {\it spine mask} (details in Sup Mat.) which allows locating individual vertebrae (Sec.~\ref{sec:anatomic_constraints}).
From the obtained locations, individual vertebrae segmentation masks
are estimated with an individual vertebrae segmentor~\citep{payer2020coarse} and refined with an iterative location-segmentation refinement scheme~\citep{lessmann2019iterative}. 
Vertebrae are further identified using the locations and segmentation masks (Sec.~\ref{sec:vert_identification}).
The obtained identifications allow, in turn, to enforce finer anatomic consistency constraints (Sec.~\ref{sec:anatomic_constraints}) and to detect possible new candidate locations.
In the latter case, the new vertebrae go through the segmentation and identification steps described before.
The process ends when the proposed consistency criteria are satisfied. It also stops when the set of detected locations, segmentation masks and identifications does not change over the cycle.
The remaining inconsistencies, such as a failure in the location-segmentation refinement or an inconsistency in the anatomic constraints, are reported in the result. 



\subsection{Anatomic consistency constraints for vertebrae localization}
\label{sec:anatomic_constraints}
Vertebrae are naturally ordered in the spine and their relative locations and consecutive sizes are heavily correlated. With the objective to exploit such prior information we compute statistics over inter-vertebral distances and individual vertebrae volumes.
Given these priors we enforce then two consistency criteria.
First, distances between all detected locations should follow the inter-vertebral distances statistics.
Second, the {\it spine mask} should be similar to the union of individual vertebrae masks.
The segmentation of a {\it spine mask} is, in general, a straightforward task which in turn allows identifying the areas where individual vertebrae should lie. In practice, at each cycle the difference between the {\it spine mask} and all individual vertebrae masks is computed to obtain residual connected components. Vertebrae volume statistics are then considered to decide whether a residual is a vertebra or just noise.
We purposely chose to use vertebrae volumes as they are more robust to morphological deviations than surface or image priors.

\textbf{The vertebrae volume constraint.}
Given a residual connected component, the idea is to check its volume by considering the volumes of the neighboring vertebrae. To this purpose a regressor is trained over consecutive vertebrae to predict the volume of the previous or next vertebrae along the spine.
The learned coefficients are in Sup Mat.

If the residual volume size is larger than a fraction of the predicted size ($50\%$ in our experiments), it is considered a vertebra and its location at its mass center is added to the vertebrae list. The residual is otherwise discarded as noise. When the vertebra identification is available, one can condition the regressors by spine levels (cervical. thoracic or lumbar), which improves the accuracy of the prediction.
Although the described volume constraint allows locating vertebrae, it should be noticed that if the {\it spine mask} presents failures, \ie on abnormal vertebrae, locations can still be missed.
It then motivates the inter-vertebral distance constraint.

\textbf{The inter-vertebral distance constraint.}
This constraint builds on the the fact that distances between locations are well structured. To this aim, we use two statistical models of the distance between vertebrae. 
The first is a Gaussian distribution for each anatomic group capturing the mean and variance of the distances between consecutive vertebrae. It is used as a prior to detect abnormally large distances.
In addition, linear regressors are used to predict the inter-vertebral distance from the neighbouring ones. In practice they predict the inter-vertebral distance from: i) the previous vertebrae distance, ii) the next one and iii) using both sides. These regressors are also conditioned on each anatomic group. Details can be found in Sup Mat. When a larger inter-vertebral distance is detected, new candidate locations are added in between. The number of new candidates is adapted depending on the current distance and the predicted one. 


We also leverage identification to check if the spine extremes are complete. If C1 (or L5) is not yet found, a location is predicted up (or down) using the two most top (down) locations. If it is inside the image field of view, it is added.


\subsection{Vertebrae identification}
\label{sec:vert_identification}
To identify vertebrae we propose to combine both a local and global reasoning.
The local reasoning is based on an individual classification network that predicts one of the $24$ possible labels for each vertebra. On top of it, the global reasoning assumes that consecutive vertebrae are {\it sorted} in the spine and aggregates the local predictions into a global, consistent identification.

\subsubsection{Individual vertebra classification.}
Our individual vertebra classification module builds on~\citep{mohammed2020morphology} where a 3D VGG network is used to classify a single vertebra into one of the $24$ categories.
The input of the network is a binary volume of size $128\times128 \times128$ that contains a vertebra segmentation mask. The output is the $N$ dimensional vector of all class probabilities. The prediction proceeds in two stages. First, a spine level, i.e., cervical, thoracic or lumbar vertebra, is predicted ($N=3$). Second, the individual vertebra goes into a per spine level classifier to predict the identity within the level ($N=7/12/5$). As shown in \citep{mohammed2020morphology}, the 
hierarchical method obtains a higher accuracy than directly classifying into one of the $24$ categories.

In our work, we extend this method \citep{mohammed2020morphology} by adding the neighboring vertebrae to the input, while keeping the target vertebra in the center of the cube, which effectively adds more context for the prediction. 
At inference time, two predictions are performed: one with the {\it spine mask} and a second with the union of all individual vertebrae segmentations. Both predictions are averaged. When an individual segmentation mask is empty, \eg  metal implants can disturb the individual segmentor, a uniform distribution over all vertebrae is used. In order to deal with transitional vertebrae, such as T13 and L6, for which only few samples appear in the training set - $1$ T13 and $13$ L6 in practice - we consider T13 as T12 and L6 as L5 and leave their fine identification to the global reasoning. 

\begin{figure}[t]
\centering
\includegraphics[width=0.45\textwidth]{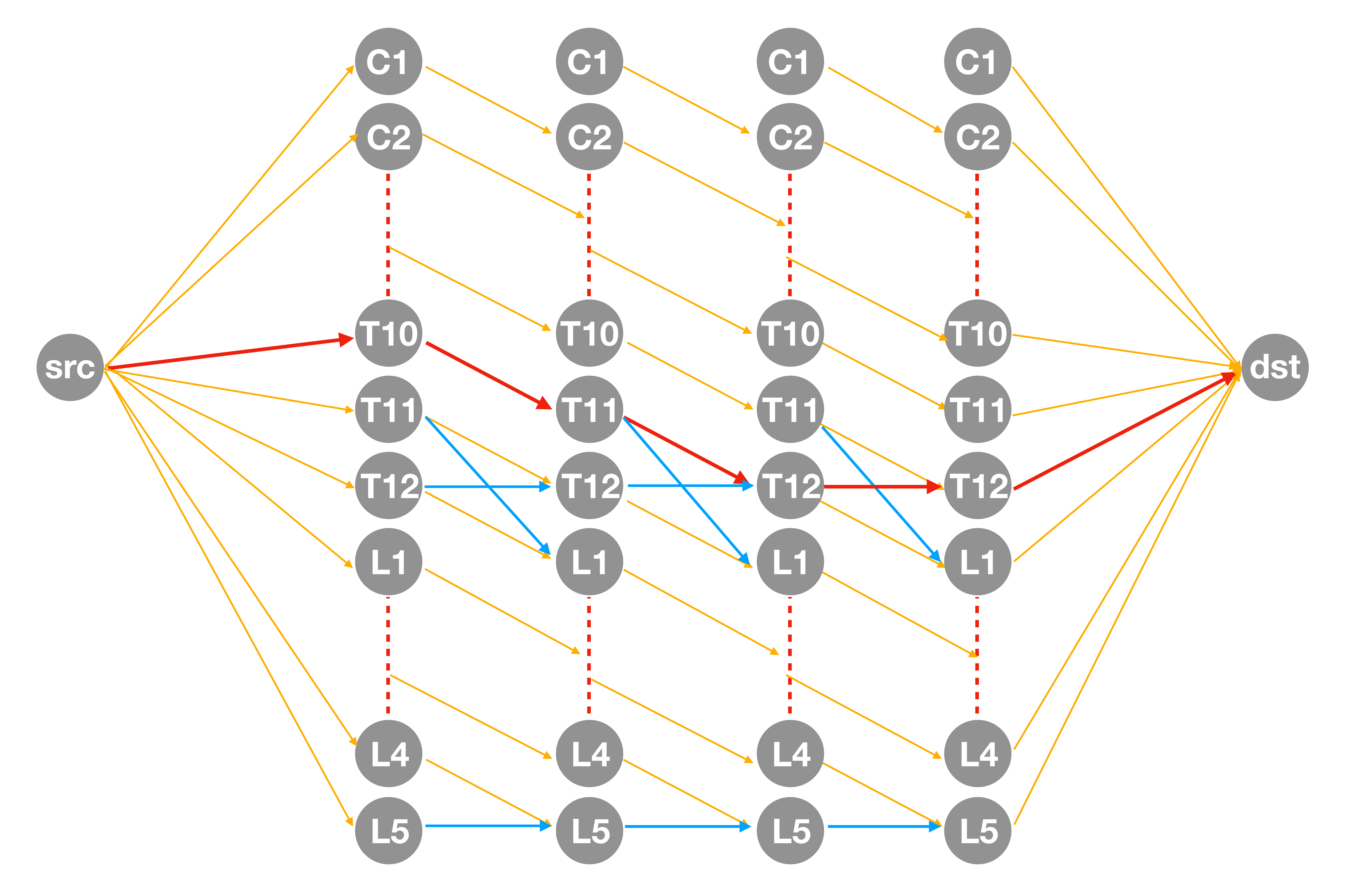}
\caption{Vertebrae identification global graph. An example with $4$ vertebrae. Edges in orange are regular connections between nodes. Edges in blue connecting T12 to T12, and L5 to L5, allow for the presence of T13 and L6. The edge between T11 and L1 allows for T12 to be absent. The result with red edges corresponds to a spine where T10, T11, T12 and T13 are observed. (Best viewed in color)}
\label{fig:graph}
\end{figure}

\subsubsection{Global graph optimization.}
Some vertebrae can still be misidentified by the local approach. In order to improve the classification, we propose a global strategy that enforces the natural ordering of vertebrae, i.e. consecutive vertebrae should have consecutive labels. We cast the problem as a shortest path search in a graph which identifies an optimal set of labels, as illustrated in Fig.~\ref{fig:graph}. Given a list with $n$ consecutive locations, a graph with $n \times 24$ nodes is created. Each of the $n$ columns represents a located vertebra, and each of the $24$ rows represents a vertebra class. Both a unary and a binary costs are then considered for the optimal path search as illustrated in Eq.~\ref{eq:unary} and Eq.~\ref{eq:binary}.
We populate the unary cost of each node with the cost obtained from the individual local probability $(1-Pv^{j}_{i})$, with $Pv^{j}_{i} \in [0,1]$. 
Additionally, as the 3-group predictions are robust, we also add a cost $Ug^{j}_{i}$ to each unary to prevent group swapping. 
The binary costs are defined on the edges between the nodes in the graph. These edges encode the fact that two consecutive vertebrae need to have consistent classes. In practice, node $N_i^j$ with $i \in [1,n]$ and $j \in [1,24]$ is only connected with $N_{i+1}^{j+1}$. 
These edges are shown in orange in Fig.~\ref{fig:graph}.
They effectively enforce that consecutive vertebrae get consecutive labels. 
Three configurations require special attention in this process, which correspond to transitional vertebrae:
i) The presence of T13, ii) The lack of T12 and  iii) The presence of L6.
To handle the presence of T13, an edge is added between two consecutive T12 nodes.
For the lack of T12, an edge is added between two consecutive nodes belonging to T11 and L1.
The presence of L6 is dealt with an edge between two consecutive L5 nodes.
These three special edges are shown in blue in Fig.~\ref{fig:graph} and are given a higher cost.
To complete the graph, two extra nodes with no cost are added to be used as the source ({\it src}) and the destination ({\it dst}) in the optimization.
The {\it src} node is connected to all nodes of the first vertebrae and the {\it dst} node is connected to all nodes of the last vertebrae.

\begin{equation}
    \mathbb{E}_{unary} = U_{src} + \sum_{i=1}^{n}{[(1-Pv^{j}_{i}) + Ug^{j}_{i}]} + U_{dst}
\label{eq:unary}
\end{equation}

\begin{equation}
    \mathbb{E}_{binary} = E_{src->n_{1}} + \sum_{i=1}^{n}{E_{i->i+1}} + E_{n->dst}
\label{eq:binary}
\end{equation}

\begin{equation}
    \mathbb{E}_{graph} = \mathbb{E}_{unary} + \mathbb{E}_{binary}
\label{eq:graph}
\end{equation}

Once the graph is created and the costs are populated, we compute the shortest path in the graph using the classical Dijsktra algorithm \citep{dijkstra1959note}.
As a post process, we check if repeated instances of T12 (or L5) are obtained and adjust the class of the second node to T13 (or L6) accordingly.

The method implementation details can be found in the released code.

\section{Experiments}

We evaluate our method using the VerSe20 MICCAI challenge dataset~\citep{loffler2020vertebral,sekuboyina2020labeling,sekuboyina2021verse}, which is currently the largest such spine dataset. It is split into a public training set, a public testset and a hidden testset. Each set consists of 100 CT scans. We follow the challenge setting using the training set data and the associated annotations to set up our method: $80$ random CT scans from the training set are used to train the networks and estimate the anatomical prior statistics, the remaining $20$ for validation.
The setting is applied throughout all experiments.


\subsection{Vertebrae identification.} 
To evaluate the benefit of using the context and the graph (Sec.~\ref{sec:vert_identification}), we compare four identification settings: (a) Using a single vertebra binary mask \citep{mohammed2020morphology}, (b) adding the neighbouring vertebrae, (c) adding the neighbouring vertebrae and using the graph. To compare to image based identification approaches we trained a classification method using the CT image as input (d). 
For a) the identification accuracy is $70.50\%$. When including the neighboring vertebrae masks, the accuracy is increased to $84.55\%$. After the graph optimization, $97.36\%$ accuracy is obtained. 
When using the CT image patch as input instead of its segmentation, a lower identification accuracy ($82.36\%$) is observed compared to the binary mask of the vertebrae ($84.55\%$). This can be attributed to the fact that the masks provide shape discriminative cues on the vertebrae.

\begin{table*}[t]
\centering
\begin{tabular}{@{}lcccccccc@{}}
\toprule
                  & \multicolumn{4}{c}{public testset}                                               & \multicolumn{4}{c}{hidden testset}                                               \\ \midrule
\multirow{2}{*}{} & \multicolumn{2}{c}{labelling results} & \multicolumn{2}{c}{segmentation results} & \multicolumn{2}{c}{labelling results} & \multicolumn{2}{c}{segmentation results} \\ \cmidrule(l){2-9} 
                  & ID rate(\%)        & MLD(mm)          & DSC(\%)             & HD(mm)             & ID rate(\%)        & MLD(mm)          & DSC(\%)             & HD(mm)             \\
Zhang A.          & 94.93              & 2.99             & 88.82               & 7.62               & 96.22              & 2.59             & 89.36               & 7.92               \\
Payer C.          & 95.06              & 2.90             & 91.65               & \textbf{5.80}      & 92.82              & 2.91             & 89.71               & \textbf{6.06}      \\
Chen D.           & 95.61              & \textbf{1.98}    & 91.72               & 6.14               & \textbf{96.58}     & \textbf{1.38}    & \textbf{91.23}      & 7.15               \\
ours              & \textbf{96.59}     & 2.21             & \textbf{92.53}      & 7.03               & 95.38              & 2.43             & 91.11               & 6.69               \\
\hline
ours(a)           & 96.02              & 2.22             & 92.06               & 6.82               & 94.93              & 2.34             & 90.85               & 6.53               \\
ours(b)           & 93.61              & 3.10             & 89.76               & 7.80               & 89.41              & 3.81             & 85.45               & 8.39               \\ \bottomrule
\end{tabular}
\caption{Quantitative comparison of our method with the top ranked methods on the VerSe20 challenge\citep{sekuboyina2021verse}
as well as with alternative (a) in which 
the anatomic consistency constraints are not conditioned on the identification and alternative (b) where the anatomic consistency priors are not used for detection.} 
\label{tb:verse_benchmark}
\end{table*}

\subsection{VerSe20 challenge results.}
We evaluate our method on the VerSe20 Challenge public and hidden testsets with 200 CT subjects in total and adopted the metrics of the challenge evaluation protocol~\citep{sekuboyina2021verse}: The \textit{Dice coefficient (DSC)} and the \textit{Hausdorff Distance (HD)} evaluate the correctly identified segmentation performance in terms of  voxel and surface similarity respectively;
The \textit{Identification rate (ID rate)} and the \textit{Mean Localization Distance (MLD)} evaluate the accuracy of the labelling task, where MLD measures the Euclidean distance of the predicted location to the GT location and $20$ mm is the criterion for a valid identification.
In the challenge metrics, when a vertebra is missed, MLD and HD are not computed. Thus there is a trade-off between ID rate and MLD (same for DSC and HD). While more positive detections lead to a better ID rate, a worse MLD can be obtained if the detections are less accurate.

Table~\ref{tb:verse_benchmark} shows the quantitative results of the proposed method and the top-scoring methods on the benchmark. Our method is on par with the best performing method in the leader-board \citep{chen2020deep} while being significantly more robust to transitional cases, as shown in Table~\ref{tb:transitional_verts}, which is our main objective. In that respect, we observe that \cite{chen2020deep} did not win the challenge, as one important criterion in the challenge was the performance in handling the transitional vertebrae - 6 cases with T13 (2/2/2 in Train/Public/Hidden), 47 cases with L6 (15/15/17) and 8 cases with absent T12(3/4/1). Following the challenge guidelines we computed the Dice score only on scans with transitional vertebrae and the obtained results are presented in Table~\ref{tb:transitional_verts}. Our method consistently outperforms all existing methods.

\begin{table}[t]
\centering
\begin{tabular}{@{}ccc@{}}
\toprule
         & public testset & hidden testset \\ \midrule
ours     & \textbf{91.04} & \textbf{89.70} \\
Payer C. & 85.96          & 89.59          \\
Zhang A. & 87.15          & 87.35          \\
Chen D.  & 84.21          & 87.01          \\\bottomrule
\end{tabular}
\caption{Methods evaluation (Dice score \%) on transitional vertebrae. \cite{sekuboyina2021verse}}
\label{tb:transitional_verts}
\end{table}

\begin{figure*}[t]
\centering
\includegraphics[width=0.8\textwidth]{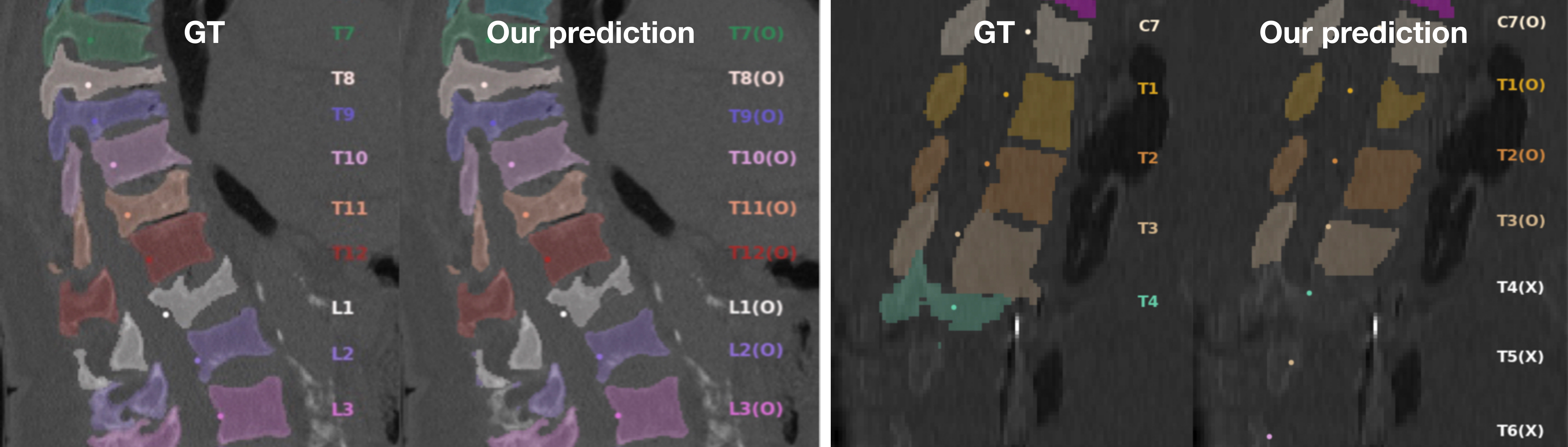}
\caption{Qualitative results on (left) fractured and (right) metal-inserted vertebrae.}

\label{fig:visual_fracture_metal}
\end{figure*}

Figure~\ref{fig:visual_fracture_metal} displays qualitative results of the method.
On the left, severely fractured vertebrae (T8,T9 and L1) are well segmented and identified. 
On the right the location-segmentation refinement \citep{lessmann2019iterative} fails due to the occurrence of metal and T4 is not segmented.
Thanks to the anatomic priors, T4, T5 and T6 can still be properly located and identified.
The cross next to the label indicates the detected inconsistency.

\begin{figure*}[t]
    \centering
    \includegraphics[width=0.8\textwidth]{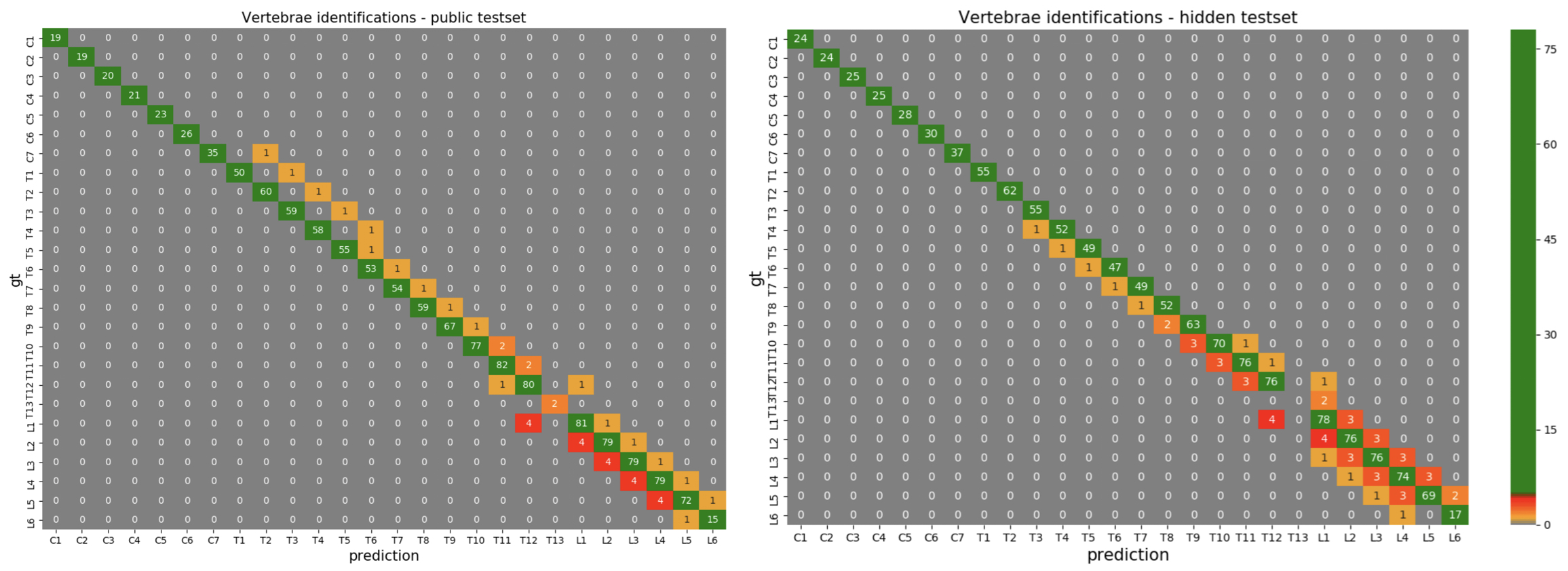}
    \caption{Final results confusion matrices. Left: public testset; Right: hidden testset.}
    \label{fig:cm_matrix}
\end{figure*}

To gain more insights into the limitations of our method, we inspected the results that obtain Dice scores lower than $90\%$ and observed that most failures are caused by wrong identification, typically one label shifts.
Figure~\ref{fig:cm_matrix} shows the confusion matrices of the final identifications. It contains $26$ classes: $24$ vertebrae from C1 to L5 and the transitional vertebrae T13 and L6. 
Most confusions arise in the thoracic-lumbar transition.
\begin{figure*}[t]
\centering
\includegraphics[width=0.8\textwidth]{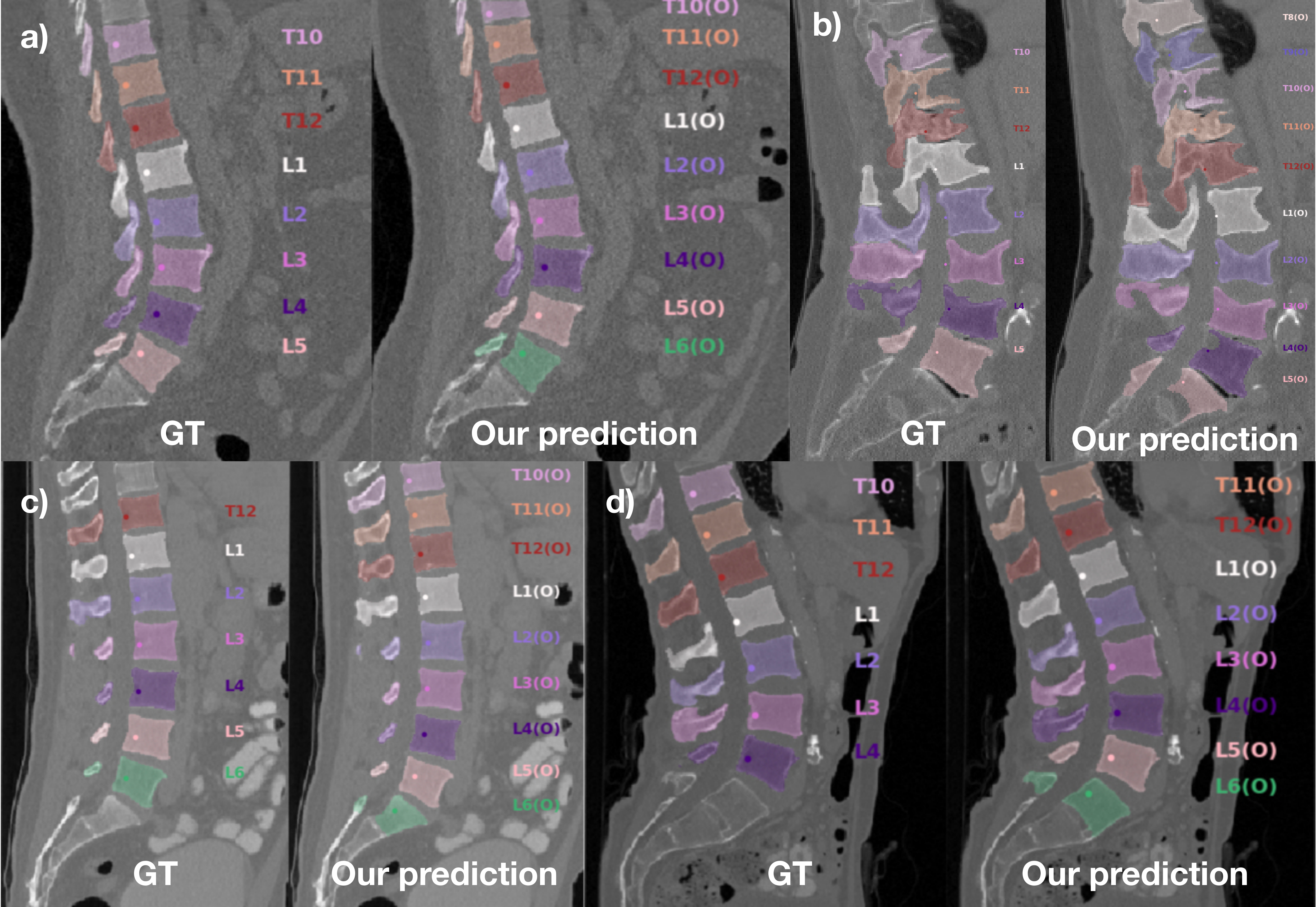}
\caption{Failure cases with one label shifted.}
\label{fig:failure_onelabelshift}
\end{figure*}

Two typical failure scenarios are observed.
First, all the vertebrae are identified with one label shifted as shown in Fig.~\ref{fig:failure_onelabelshift}.
The confusions are around thoracolumbar area. To investigate if the identification error is from the individual vertebra prediction or the graph optimization, we inspect the probabilities of the individual prediction. We found that the individual vertebra classification network holds high confidence in its outputs.
Take case c) in Fig.~\ref{fig:failure_onelabelshift} as an example, the labels from top down predicted by the network are [17, 18, 19, 20, 21, 22, 23, 24, 25]\footnote{The numerical labels 1-25 correspond to the vertebral levels C1-L6} (two more vertebrae detected than GT) with probabilities [1.0, 1.0, 1.0, 0.99, 1.0, 1.0, 1.0, 0.99, 1.0], while the GT labels are [19, 20, 21, 22, 23, 24, 25]. As we know, one of the challenges of vertebrae identification is the high similarity of the neighboring vertebrae shapes that would confuse the deep neural network. On the contrary, the network is much confident about its prediction without any uncertainty between options. This implies that either the network is not capable enough to extract the implicit features of each vertebra, or the network is not capable to classify and differentiate them. The drawback points to the network architecture design, where we use a baseline backbone architecture \textit{3D vgg16} (Sec.~\ref{sec:vert_identification}). Further improvements can be made towards the optimization of the individual vertebra classification network. 

\begin{figure*}[t]
\centering
\includegraphics[width=0.8\textwidth]{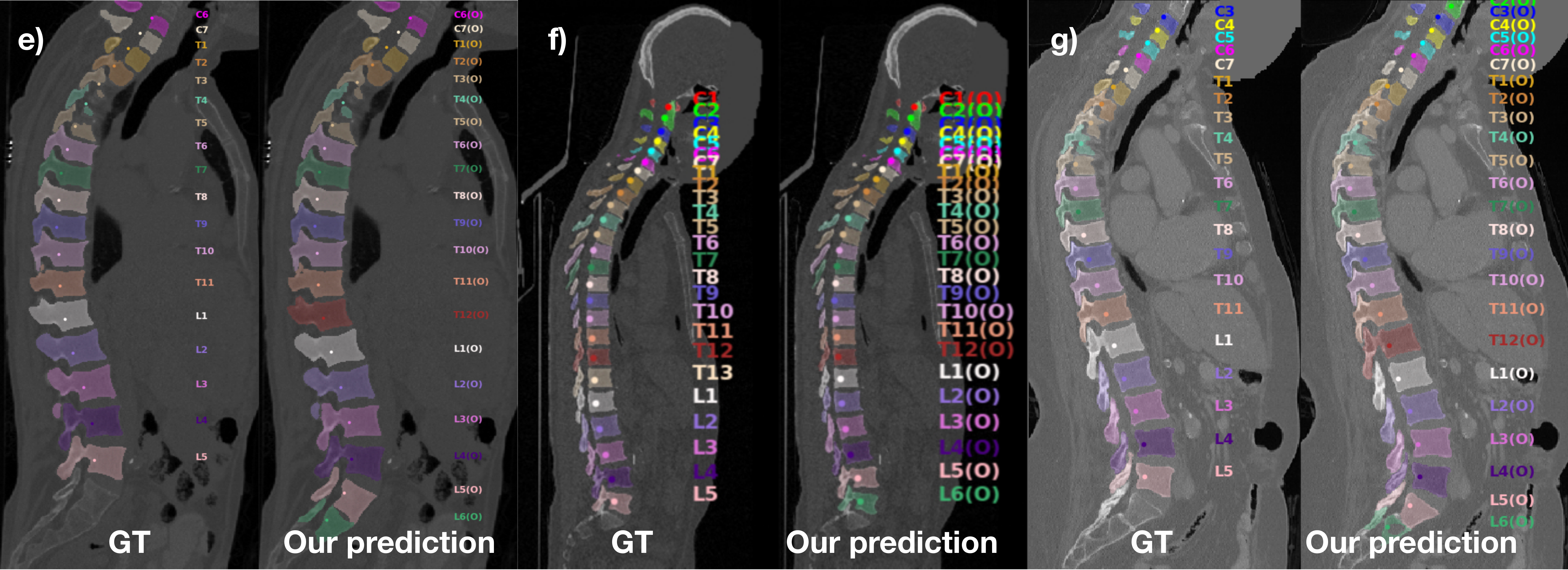}
\caption{Failure cases on transitional vertebrae.}
\label{fig:failure_transitional}
\end{figure*}
Second, the identification fails on challenging transitional vertebra and the consequent labels are shifted, as shown in Fig.~\ref{fig:failure_transitional}. 
The graph explicitly models the transitional vertebrae. Nevertheless, it is still a challenging task. One possible reason for the failure is the imbalance of the presence of the transitional vertebrae in the training and test sets. Transitional vertebrae are rare in the dataset. Moreover they are unproportionally distributed in the training and test sets to challenge the algorithm. In our method, the cost of bridging the abnormal transitions is higher and weighted, where the weights for the three transitions (presence of T13, L6 and the absence of T12) are turned with the training set. It does not generalize the best performance to the test sets. The improvement can be explored in learning the weights for the transitional vertebrae and the other individual vertebrae.

\subsection{Ablation study.}
To highlight the benefits of the different parts of the proposed method we perform two experiments.
In the first one ({\it ours(a)} in Tab.~\ref{tb:verse_benchmark}) we do not use the identification to condition the vertebrae statistics. 
Instead of learning specialized regressors for each spine level, we use a fixed threshold to determine if a residual mask is a potential true vertebra (we use half of the smallest vertebra volume size in the training set: $7820 / 2$).
To decide if there is a missing vertebra given a distance between $2$ we use a fixed threshold of $50$ mm \citep{payer2020coarse}. The overall scores show that the improvement when using conditional statistics is minor but not negligible: without conditioning $8$ vertebrae are missed in the public testset and $6$ in the hidden test set while with the identification information no vertebra is missed in public testset and only $4$ in hidden testset. 
In the second experiment ({\it ours(b)} in Tab.~\ref{tb:verse_benchmark}) the anatomic consistency priors (Sec.~\ref{sec:anatomic_constraints}) are not used. Without the constraints on vertebrae volume and the relative locations 
the performance significantly drops.

\section{Generalization and limitation}
We further evaluate our method on other spine datasets to investigate the generalization capability and the limitation of the method. These datasets include the publicly available datasets with associated benchmark where we can compare with the state-of-the-art approaches, and in-house datasets. We provide both quantitative and qualitative results.

\begin{table*}[t]
\centering
\scalebox{0.85}{
\begin{tabular}{@{}cccccc@{}}
\toprule
Dataset                     & Method            & Dice(\%) & ID rate(\%) & ASSD(mm) & HD(mm)     \\ \midrule
\multirow{4}{*}{VerSe19}    & ours              & 90.84    & -           & -        & -          \\
                            & Chen D.\textsuperscript{\#}          & 86.44    & -           & -        & -          \\
                            & Payer C.\textsuperscript{\#}           & 84.11    & -           & -        & -          \\
                            & Zhang A.\textsuperscript{\#}          & 85.42    & -           & -        & -          \\ \midrule
\multirow{5}{*}{LumbarSeg}  & ours              & 95.3$\pm$0.3 & 90          & 0.3$\pm$0.1  & 7.4$\pm$1.5    \\
                            & \cite{lessmann2019iterative}\textsuperscript{\#}     & 96.5$\pm$0.8 & 100         & 0.2$\pm$0.0  & -          \\
                            & \cite{korez2015framework}      & 95.3$\pm$1.4 & -           & 0.3$\pm$0.1  & -          \\
                            & \cite{chu2015fully}        & 91.0$\pm$7.0 & -           & 0.9$\pm$0.3  & 7.3$\pm$2.2    \\
                            & \cite{ibragimov2014shape}  & 93.6$\pm$1.1 & -           & 0.8$\pm$0.1  & -          \\ \midrule
\multirow{6}{*}{xVertSeg}   & ours              & 94.6$\pm$0.9 & 100         & 0.6$\pm$0.1  & 13.06$\pm$6.71 \\
                            & \cite{lessmann2019iterative}\textsuperscript{\#}    & 94.6$\pm$2.2 & 100         & 0.3$\pm$0.2  & -          \\
                            & \cite{janssens2018fully}   & 95.7$\pm$0.8 & 100         & 0.4$\pm$0.1  & 4.32$\pm$2.60  \\
                            & \cite{sekuboyina2017localisation} & 94.3$\pm$2.8 & -           & -        & -          \\
                            & \cite{chuang2019efficient}\textsuperscript{\#}      & 88.5     & 100         & -        & -          \\
                            & \cite{cheng2021automatic}      & 87.7$\pm$3.5 & 100         & -        & -          \\ \midrule
\multirow{8}{*}{CSI2014Seg} & ours              & 93.4$\pm$0.8 & -           & 0.7$\pm$0.1  & 6.6$\pm$1.2    \\
                            & \cite{kolavrik2019optimized}   & 97.1$\pm$0.0 & -           & 0.3$\pm$0.1  & -          \\
                            & \cite{lessmann2019iterative}\textsuperscript{\#}  & 96.3$\pm$1.3 & 81          & 0.1$\pm$0.1  & -          \\
                            & \cite{lessmann2018iterative}\textsuperscript{\#}  & 94.8$\pm$1.6 & -           & 0.3$\pm$0.1  & -          \\
                            & \cite{cheng2021automatic}\textsuperscript{\#}      & 95.3$\pm$1.4 & -           & -        & 4.0$\pm$2.1    \\
                            & \cite{korez2015framework}     & 94.6$\pm$2.0 & -           & 0.3$\pm$0.1  & -          \\
                            & \cite{korez2015interpolation}     & 93.1$\pm$2.0 & -           & -        & -          \\
                            & \cite{hammernik2015vertebrae}  & 93.0$\pm$4.0 & -           & -        & -          \\ \midrule
\multicolumn{6}{l}{{\#} same training/evaluation settings as ours} 
\end{tabular}
}
\caption{Quantitative results of the state-the-of-art methods on publicly available spine CT datasets.}
\label{tb:other_spine_data}
\end{table*}

\paragraph{VerSe19 dataset}
The VerSe19 dataset~\citep{sekuboyina2021verse}, released as part of the  VerSe challenge in MICCAI 2019, consists of $80$ training CT scans with the associated annotation and $40$ testing CT scans for public and hidden testsets respectively. The dataset is rich with varying field of views.
To evaluate the generalization capability of our method, we perform the experiment of the Verse20 challenge~\citep{sekuboyina2021verse} where the approach developed and trained on the VerSe20 training set is applied to the VerSe19 hidden testset. Table~\ref{tb:other_spine_data} shows that our method outperforms all approaches and generalizes well to VerSe19 testset. 

\paragraph{Lumbar vertebra segmentation CT image database}
The LumbarSeg dataset consists of $10$ scans of healthy subjects and associated annotation for lumbar vertebrae. The slice resolution is between $0.28$ mm to $0.79$ mm and the slice thickness is between $0.72$ mm to $1.53$ mm.
We follow the evaluation protocol of \cite{lessmann2019iterative} and use it for an external evaluation of our supervised method. Scans from this dataset were only used for evaluation and were not part of the training set.
For the segmentation we obtain Dice score (\%) of $95.3\pm0.3$ which is consistent with the performance on VerSe20 dataset.
For identification accuracy, we obtained $45$ correctness out of $50$ vertebrae. One subject failed on the \textit{one-label-shift} issue. The L5 was recognized as L6 and the rest vertebrae above it are with one label offset. This failing scenario is also found in VerSe20 dataset evaluation.
For other methods, \cite{korez2015framework} achieved on-par performance when using a different evaluation protocol, where they split the 10 CT subjects into train and test sets. 
\cite{chu2015fully} evaluated their method with a leave-one-out cross validation. Only vertebral bodies were segmented and evaluated.
Quantitative results can be found in Table~\ref{tb:other_spine_data}.

\paragraph{xVertSeg dataset}
The xVertSeg dataset \citep{ibragimov2017segmentation}, released as part of the xVertSeg challenge in MICCAI 2016, consists of $15$ train CT volumes with ground truth segmentation of the lumbar vertebrae (into five classes, L1-L5) and $10$ test CT volumes. The ground truth segmentation of the test set is not publicly available. The dataset includes non-fractured vertebrae and vertebrae with fractures of different morphological grades and cases. The scans were reconstructed to in-plane resolutions of $0.29$ mm to $0.80$ mm and slice thickness of $1.0$ mm to $1.9$ mm. The scans are with varying field of views, while mostly observing the lumbar area.
There are methods that are evaluated using the same dataset, but with different evaluation/training protocols. 
\cite{janssens2018fully} conducted a leave-three-out cross-validation study to evaluate the performance of their proposed method. More specifically, each time they randomly took $3$ out of $15$ CT data as test data and the remaining  $12$ CT data as the training data. The process was repeated $5$ folds.
\cite{sekuboyina2017localisation} used the $15$ training set to train their model and evaluate the performance on test data. As the ground truth annotation of the test data is not released, they opted for an in-house ground truth generation. The near-perfect segmentation from their approach was given to two clinical experts (Rater-1 and Rater-2) for correction. Rater-1 was tasked to correct the entire volume, while Rater-2 was tasked to pick a random subset of sagittal, coronal and axial slices from a volume and segment them entirely. Using the ground truth generated from Rater-1, they obtained $94.3\pm2.8$ Dice accuracy (\%). $92.0\pm2.3$ Dice score (\%) was obtained when using Rater-2 ground truth segmentation.
\cite{cheng2021automatic} divided the $15$ training data into two parts: $10$ images for training and $5$ for testing. The split information was not explicitly provided. 
To conduct a comparative evaluation, we use the scans $1$ to $5$ for evaluation and the remaining $10$ scans for training, which is the same setting introduced in \citep{lessmann2019iterative,chuang2019efficient}. Therefore, we can compare our method to theirs. Quantitative results are shown in Table~\ref{tb:other_spine_data}.

\paragraph{CSI2014 segmentation dataset}
The CSI2014Seg dataset, released as part of the spine segmentation challenge in MICCAI 2014, consists of 15 CT scans of healthy subjects. Full thoracic and lumbar vertebrae (17 vertebrae in total) are observed and their corresponding segmentations are provided. The scans were reconstructed to in-plane resolutions of $0.31$ mm to $0.36$ mm and slice thickness of $0.7$ mm to $1.0$ mm. The dataset is split into training set of $10$ scans and test set of $5$ scans. We evaluate our method following the challenge train/test split, same as \cite{lessmann2018iterative,lessmann2019iterative} and Cheng \cite{cheng2021automatic}.
Only the $10$ training data were used for other approaches.
\cite{hammernik2015vertebrae} and \cite{korez2015framework,korez2015interpolation} performed a leave-one-out 10-fold cross validation and reported the average accuracy over 10 experiments. 
\cite{kolavrik2019optimized} evaluated their approach using a leave-one-out 3-fold cross validation.

\paragraph{EOS imaging test data}
We received $3$ test data from EOS imaging, France.
All the three CTs are scanning the full spine from C1 to L5.
The first two subjects are elder patients, both with contrast agents injected in the spinal cords. For Dice scores (\%) we obtain $85.88\pm3.11$ and $91.05\pm1.68$ respectively. For HD (mm) we obtain $7.33\pm2.01$ and $6.30\pm2.69$ respectively.
The third subject is a young adult with scoliosis. We obtain Dice score (\%) of $93.38\pm2.00$ and HD (mm) of $4.45\pm4.86$.
The obtained accuracy indicates that our method generalizes well to unseen datasets and provides adequate segmentation and identification.

\begin{figure}[]
\centering
\subfigure[Subject with scoliosis]{\includegraphics[width=0.22\textwidth]{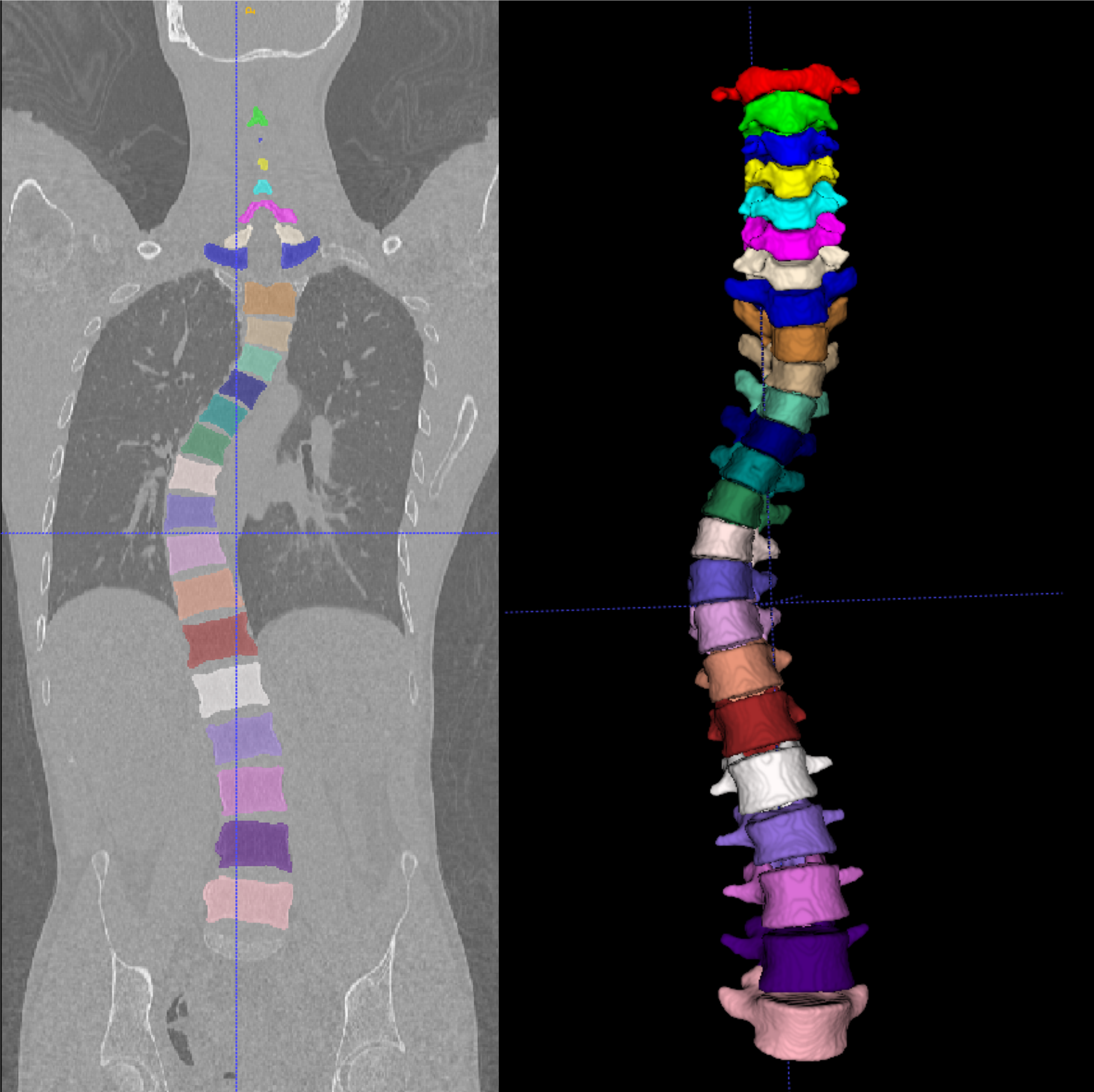}}
\subfigure[Subject with contrast agents]{\includegraphics[width=0.23\textwidth]{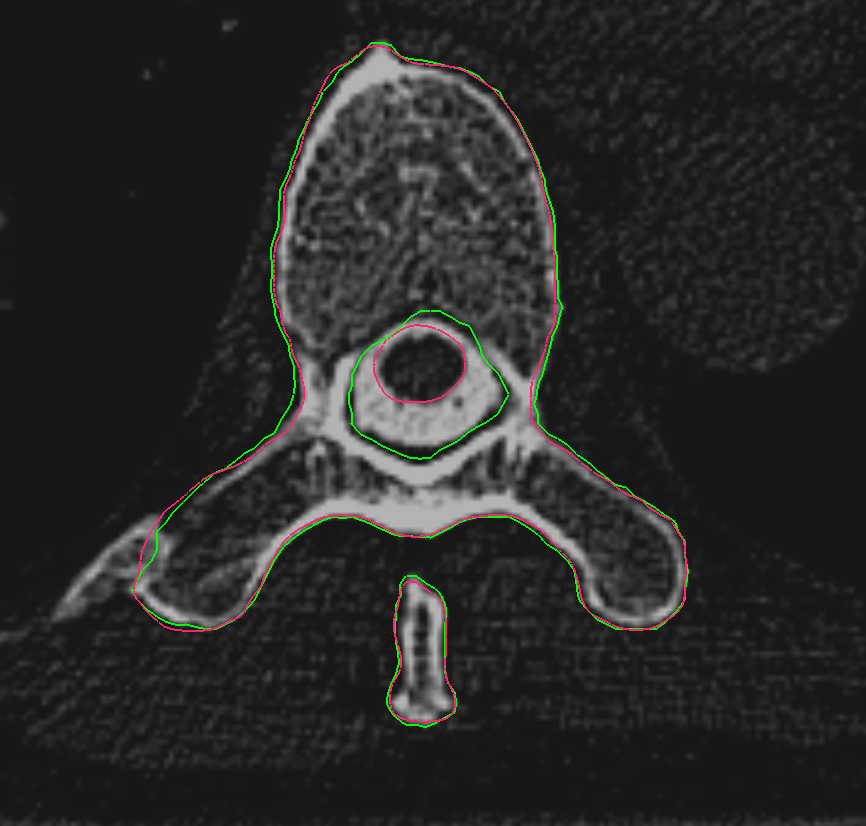}}
\caption{Results visualization of special cases; (a) Vertebrae segmentation and identification of a full spine with scoliosis; (b) An axial slice of the vertebra. Green line is the GT segmentation, red line is our prediction. The contrast agent is segmented as bone by our method.} 
\label{fig:eos_limitation}
\end{figure}

By inspecting the results slice by slice, we found that the contrast agents injected in the patients' spinal cord are segmented as bones. As shown in Figure~\ref{fig:eos_limitation}~(b), the green and red lines represent the contour of GT segmentation and our prediction respectively. The brightness caused by the contrast agent presents different and new features that the training set did not observe, suggesting that our method is not robust to this contrast agent.
Nevertheless, our method is well generalized to the scoliotic spine. Figure~\ref{fig:eos_limitation}~(a) shows the segmented spine which presents a moderate curve ($25-40$ degrees). By virtue of the data augmentation strategy, our method is robust to the tilted vertebrae.

\paragraph{CHU test data}
This test set is from CHU Grenoble, France. We are blind to this test set in the way that the hospital took our packed implementation and tested our method on their own. We have no access to the test data and only received the statistics of the results from them.
The test data consists of $675$ CT scans associated with the manual annotations from the doctor. All the scans include the abdomino-pelvic cavity, where part of the spine is presented,  but are not restricted to this field of view. 

\begin{table*}[]
\centering
\begin{tabular}{@{}ccc@{}}
\toprule
Dice score (\%)                 & with the initial annotations & with the refined annotations \\ \midrule
including the empty predictions & 89.9                         & 93.6                         \\
excluding the empty predictions & 92.8                         & \textbf{96.5}                \\ \bottomrule
\end{tabular}
\caption{Quantitative results of CHU test set.}
\label{tb:chu_testset}
\end{table*}

Among the $675$ CT scans, $20$ scans did not provide any output because of the memory issue or non-converged individual vertebra segmentor. The rest provided promising results where the accuracy can be found in Table~\ref{tb:chu_testset}.
After obtaining the vertebrae segmentation and identification by running our code, the doctor further refined their initial manual annotation taking our prediction as a reference. We see that with the refined ground truth, we obtain $93.6\%$ Dice score, $3.7\%$ higher than that with the initial annotation. Excluding the $20$ invalid outputs, the Dice score can reach $96.5\%$.

A significant fact was reported that the doctor corrected their initial annotation by adding $25$ L6. They are missed in the manual segmentation but detected by our method. It indicates the normality of sacralization/lumbarization in the population. In total around $10\%$ L6 are found in the test set which is coherent with the statistics \citep{carrino2011effect,uccar2013retrospective,konin2010lumbosacral}.

Apart from the $20$ scans that provided empty output, among the $675$ test scans, $23$ scans obtained unqualified predictions.
Most of them have errors in the L5/Sacrum area. Our method tends to look for L6 in the field of view, some S1 which are not separated from the sacrum are being well segmented and identified as L6.

\begin{figure}[]
\centering
\subfigure[Spine with scoliosis]{\includegraphics[width=0.21\textwidth]{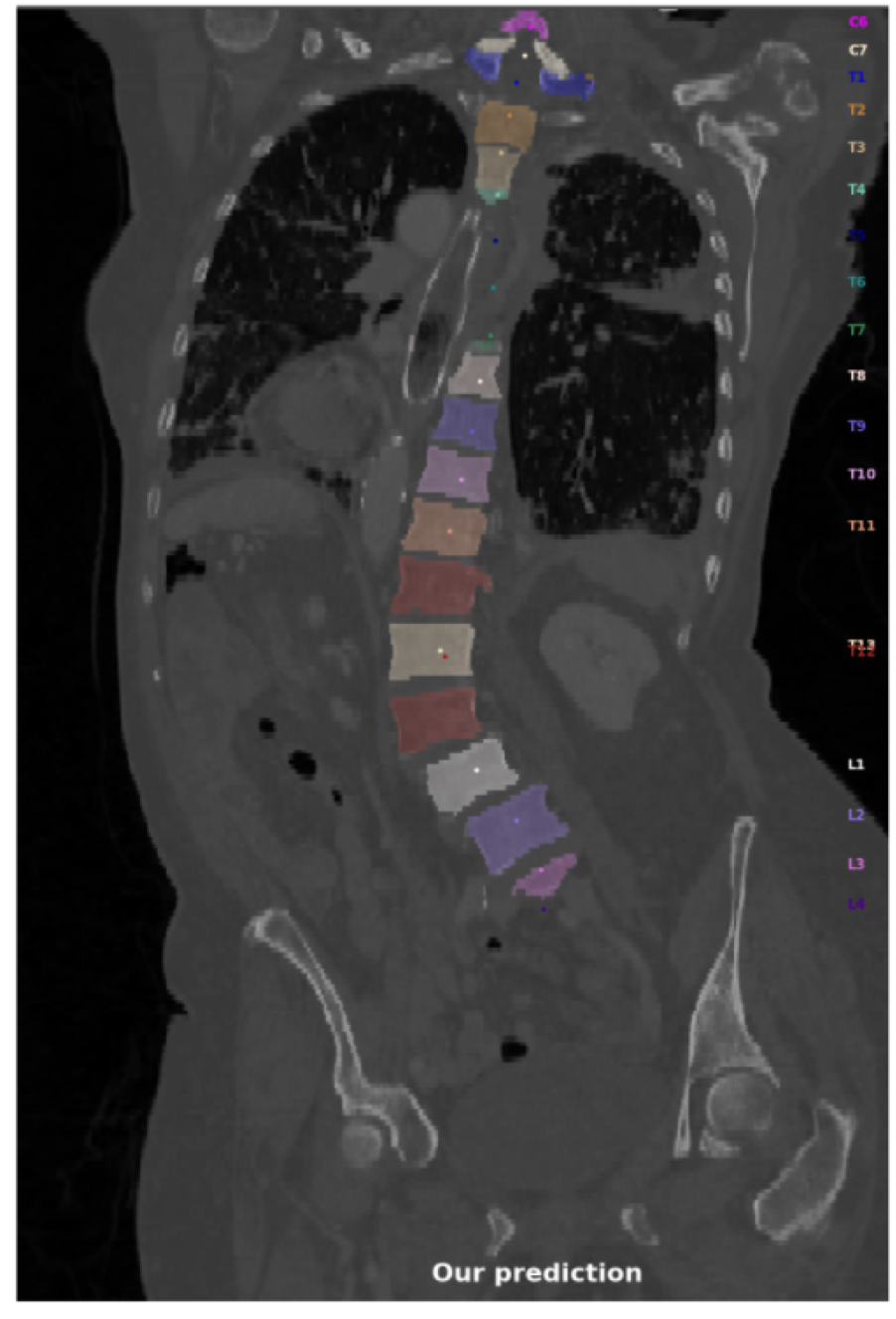}}
\subfigure[Spine with kyphosis]{\includegraphics[width=0.232\textwidth]{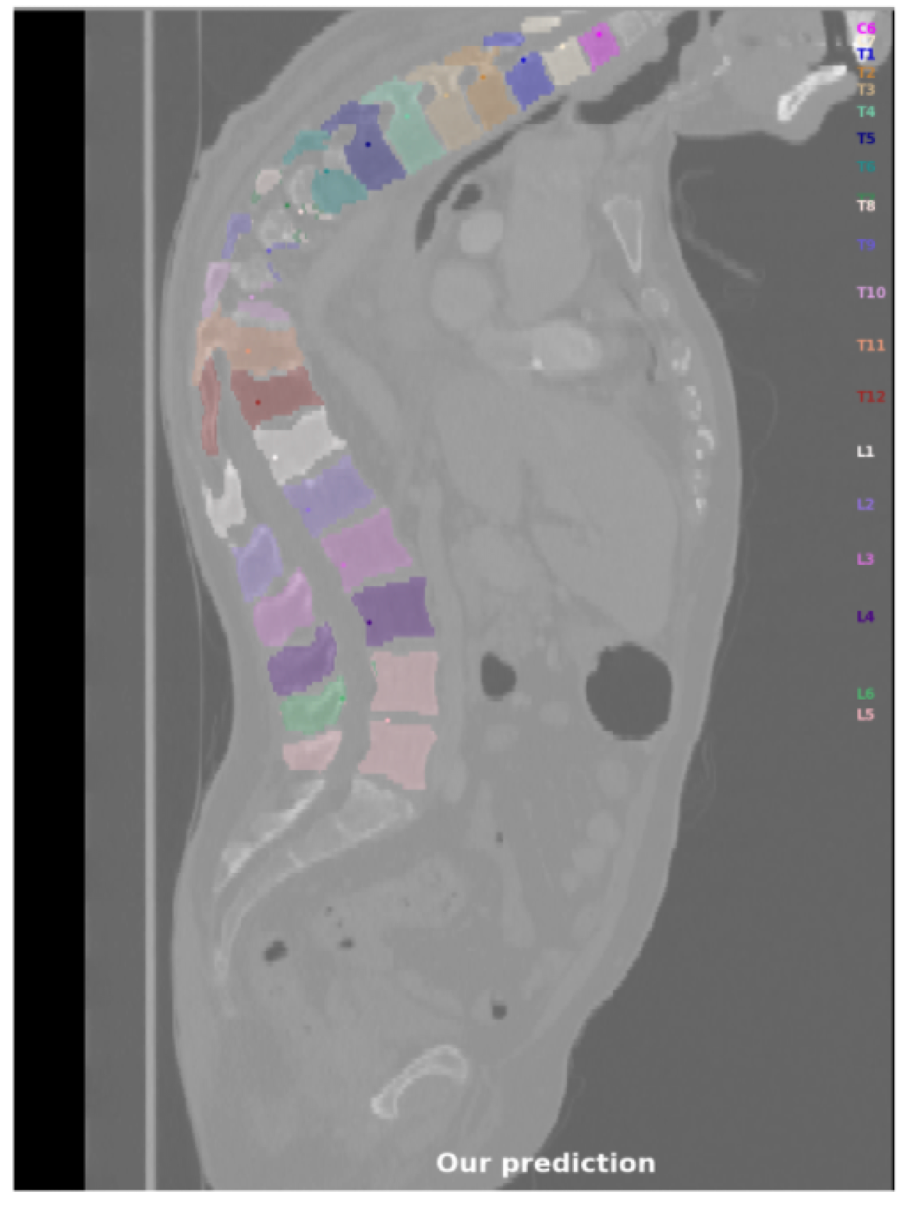}}
\caption{Failures on severe scoliosis and kyphosis.}
\label{fig:chu_scoliosis}
\end{figure}

A few failure cases are observed in abnormal subjects or the ones with severe deformities, such as severe scoliosis and kyphosis as shown in Fig.~\ref{fig:chu_scoliosis}. The spine with severe kyphosis has up to $90$ degrees curvature, which means the axial plane of the vertebrae is orthogonal to the axial plane of the CT volume. In our algorithm, we pre-process all the input scans into the same anatomical orientation. The spine is allowed to have some rotations ($50$ degrees tolerance) as we augmented the training data. The trained network is not robust to the severe scoliosis where the vertebrae are largely rotated. To lift this limitation, future work can augment the training data with larger variance. 

For the failure cases of the non-converged individual vertebra segmentation, it either segments nothing at the detected location, or it segments something but loses it in the iterative location-segmentation refinement scheme. A hypothesis is that the acquisition parameters of the CT are set differently for these subjects, for specific-area diagnosis purposes. Thus, the intensity and the contrast of the image are affected and not distributed as the training set.

\begin{figure*}[]
\centering
\subfigure[Scan with metals]{\includegraphics[width=0.32\textwidth]{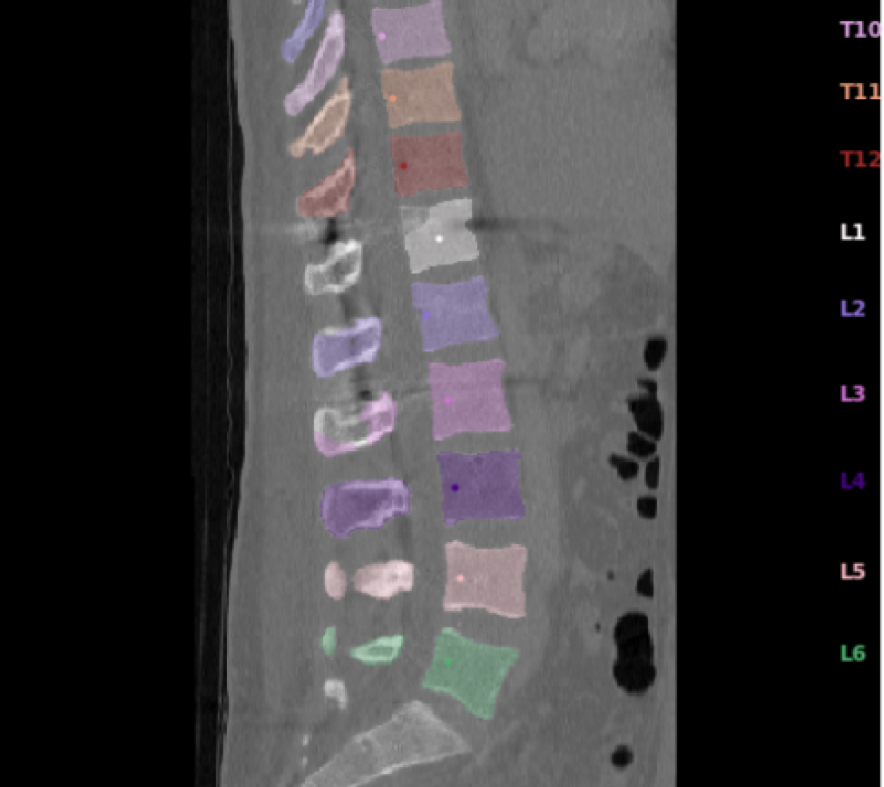}}
\subfigure[Scan with poor quality]{\includegraphics[width=0.253\textwidth]{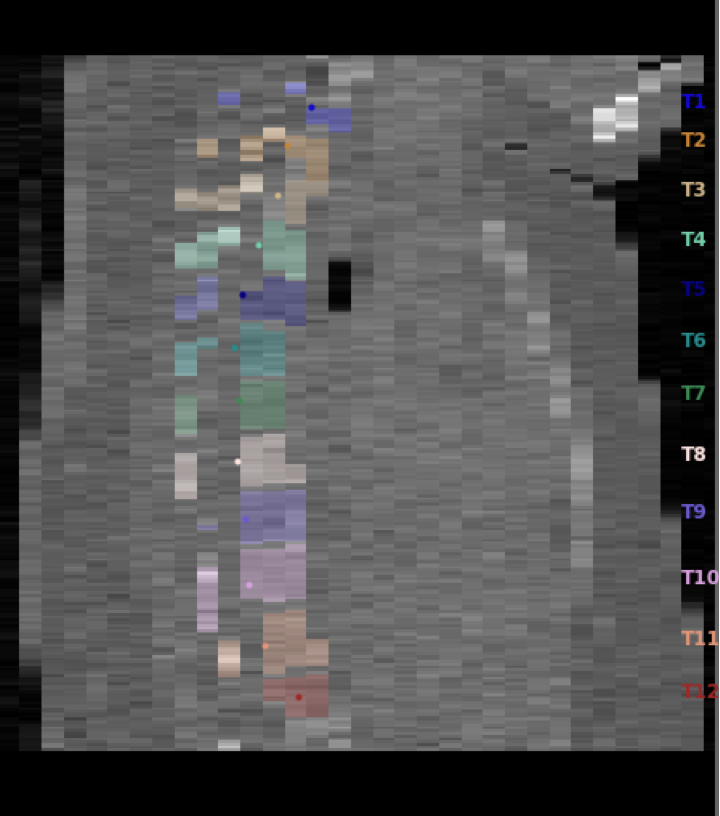}}
\subfigure[Scan with fractured vertebrae]{\includegraphics[width=0.31\textwidth]{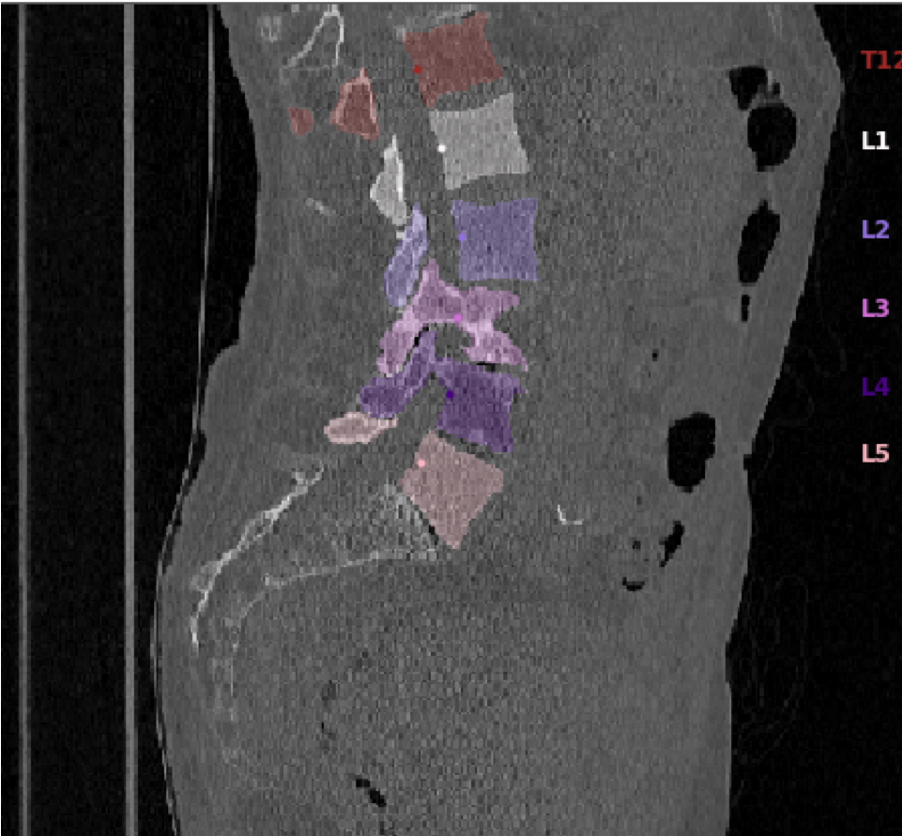}}
\caption{Qualitative results on extreme cases.}
\label{fig:chu_visu}
\end{figure*}

Overall, our method achieved considerable results on this large test set. For some extreme cases such as metal-inserted vertebrae, fractured vertebrae and very poor-quality scans, it also provided promising outputs as shown in Fig~\ref{fig:chu_visu}.

\section{Discussion}
We presented an anatomic consistency cycle for vertebrae localization, segmentation and identification in arbitrary field of view CT images. The set of vertebrae locations, masks and labels is cycled through until an anatomical coherence is achieved. Compared with most existing methods, which use a chain of modules to perform the three tasks, the cycling process avoids the accumulated errors from the sequential blocks. Moreover, the three tasks cooperate and complement each other boosting the performance. In our method, the vertebra location and segmentation benefit and refine each other. The anchor location is used to segment the vertebra and it can in turn be computed from the segmented mask. A stable pair of vertebra location and segmentation is obtained through the iterative process. The iterative refinement scheme is inspired by \cite{lessmann2019iterative}, in their method a segmentation patch guides a sliding window until a complete vertebra is seen. The vertebra location and its corresponding mask are then used for identification. The identification in turn conditions on the vertebral statistics to locate individual vertebrae. The cycle unifies the three tasks and utilizes the anatomy, assuring the obtained set of locations, segmented masks and labels is anatomically consistent.
We regard the proposed anatomic consistency cycle as a general framework for multi-label segmentation task in medical images, in the sense that the individual modules can be replaced. For instance, the segmentation or the identification network can be customized \citep{payer2020coarse,tao2022spine}. We use the baseline networks such as U-net \citep{ronneberger2015u} and vgg \citep{simonyan2014very}, as designing a new architecture is beyond the scope of this work. The framework can be generalized to other parts of the body, \eg ribs, as the statistics of the anatomy can be encoded and learned. Additionally, the cycle with the defined consistency criteria can be used as a post-processing step to refine the result.

In the consistency cycle, we proposed to leverage the anatomical priors with deep neural networks. Generally a large amount of data benefit the supervised learning, hence low present samples are difficult to be learnt. In this work, we combine both strategies where the deep networks promise the learned outcome while the anatomical priors save the abnormal cases, \eg pathological or transitional vertebrae.
Fractured, compressed or other pathological vertebrae present non-standard shapes, thus the hard constraints such as the appearance and image models do not apply.
We chose to learn the statistics of the vertebra volume and the inter-vertebral distance. The volume quantifies a part of the vertebra shape. It is robust to morphological deviations. The inter-vertebral distance regularizes the vertebrae distributions. The two features are used as soft constraints to locate the individual vertebrae.
The located vertebrae are then segmented by the individual segmentor. Through the iterative refinement process, either the location is stable with the segmentation mask and added, or the mask is empty and the location is discarded. In the circumstance that the location added by the anatomic constraints is not consistent with the segmentor, we give the anatomy the priority and add the location. The hypothesis for this inconsistency is that the individual vertebra patch is highly distorted, \eg there is a metal implant. 
The anatomical priors are additionally used as consistency criteria validating that the final output is anatomically coherent. We report the inconsistent region in the result, available for the specialist to further diagnose.

We use a hierarchical approach for the individual vertebra classification task. A vertebra is classified through two stages, first the anatomic group is identified and then the individual label. Similar strategies are also found in \citep{sekuboyina2017localisation,masuzawa2020automatic} where they first separate the spine into different anatomic regions, then segment and label the individual vertebrae. The two-stage method largely reduces the intra-class variances and improves the prediction accuracy.
We employed the vertebrae morphology for the classification. Most existing works use the CT image as input to classify vertebrae with the intuition that the original scan provided more information. Our motivation was to investigate if the shape of the vertebrae itself can be distinguished. We experimentally showed that the vertebrae shape is representable compared with the image patch, higher/comparable accuracy was achieved. The advantage of classifying vertebra from its shape is modality independent. The trained classifier is generalized and not restricted to one modality (\eg CT or MRI) particularly. However, the prerequisite of using the shape is to segment the spine from the original scan. Vertebrae with incomplete or corrupted shape would thus affect the precision of the prediction. Even with the completely segmented vertebrae, the individual predictions may have errors. It motivated us to combine the local predictions with a global reasoning. The proposed graph enforces the consistency over individual predictions. It tolerates a certain number of inaccurate individual predictions and sorts out an optimal configuration. 
Another key feature of the proposed graph is that it explicitly models the transitional vertebrae, which affect $\sim25\%$ of the general population but are limited present in the publicly available training set. With the graph, our method achieved state-of-the-art performance in detecting the transitional vertebrae. It is worth noting that in the evaluation of the in-house dataset, $25$ extra L6 were detected among $675$ CT scans. Our method well handles the corner cases while maintaining a good performance on general data. 

Overall, our method achieved the state-of-the-art performance on VerSe challenge benchmark and obtained comparable results on other publicly available datasets. On the unseen in-house datasets, our approach generalizes well and provided promising results. We experimentally showed that the method is robust to fractured, metal-inserted and compressed vertebrae, mild to moderate scoliosis, extreme poor-quality scans. Yet, it still finds challenges in severe scoliotic spine and transitional vertebrae. A typical failing situation of the method is the one-label-shift issue. We found that with the limited field of view, usually when only the thoracolumbar area is observed, the individual vertebra classification provides high confidence in predicting one-offset labels. We use a baseline architecture vgg for the classification. Further improvements can be realized on optimizing the classification network or data augmentation.

Considering the time and GPU cost of our method, inference times on VerSe20 public testset ($103$ CT scans) vary between $1.5$ min to $248$ min depending on the scan size, with a median runtime of $26$ min.
For the in-house dataset of $675$ CT scans, the inference time varies between $7-90$ min ,with a median runtime of $16.2$ min. The time is heavily devoted to the spine binary segmentation, as the spine is segmented with intensely overlapped patches. It can be drastically improved by considering parallelization over patches. It also costs a lot when there are more than two loops in the anatomic consistency cycle.
Nevertheless, the method is GPU efficient with no more than $3$ GB occupation. The CPU memory is costly when the individual vertebrae volumes are processed and accumulated in the cycle.

\section{Conclusion}
In this work we presented a new approach to localize, segment and identify vertebrae in CT images. The 
approach combines deep-learning networks with statistical consistency priors on vertebrae, improving robustness to transitional vertebrae and pathological cases. An anatomic consistency cycle is proposed that aggregates task specific deep-networks while enforcing statistical priors.  The evaluation on the standard VerSe20 challenge benchmark demonstrates the interest of the proposed strategy, in particular with transitional vertebrae which are present in a meaningful part of the population. The extension of the proposed anatomic consistency cycle to other anatomic structures with similar specific cases is currently under investigation. 

\section*{Acknowledgments}
This work was jointly supported by the Fonds Unique Interministériel [grant number 1701595901-N00013834], the Auvergne Rhône Alpes region and the Grenoble Alpes metropole. The networks training in this work was performed using HPC resources from GENCI-IDRIS (Grant 2021-[AD011012208]). We thank Dr. Kazu Hasegawa for the test dataset and the EOS imaging team for the manual expert segmentations used in the assessment. We thank Charlene Dumont for testing the code and the CHU Grenoble for the test dataset with manual expert segmentations.

\bibliography{refs.bib}

\section*{Supplementary Material}
\subsection*{A. Spine and vertebra segmentation}

\paragraph{Segmentation network.}
To segment bones (i.e. the full spine or individual vertebrae) we use a 3D Attention U-net. The backbone architecture is a 3D U-net \citep{ronneberger2015u} with attention blocks \citep{oktay2018attention} embedded in the skip connections. The network input is a 3D CT image patch and the output is a probability mask for each voxel indicating if it belongs to the bone or not.

To train the network we use a regression loss $\mathbf{L}$ that combines the Dice coefficient and the Mean Squared Error, the objective being to minimize the difference between our prediction and the ground truth. 
We note the input $x$, the output $G(x)$, the ground truth patch as $y$ and define

\begin{equation}
\mathbf{L}(G(x), y) =1-\frac{2\lvert G(x)\cdot y \rvert}{\lvert G(x) \rvert ^2 + \lvert y \rvert^2}  + \lambda \|y-G(x)\|_2,
\label{loss:dice_l2}
\end{equation}

\paragraph{Spine segmentation.}
As the input CT has an arbitrary size and the network accepts only a fixed size input, we use patches of size $96 \times 96 \times 96$. In Eq.~\ref{loss:dice_l2} $\lambda$ is empirically set to $10$ and kept constant during the training. At inference time, the trained model is applied in a sliding mode over the whole CT volume with a stride length of $24$ (one quarter of the patch size) and for each voxel we obtain 64 predictions. We aggregate the results in each voxel by averaging the $64$ probabilities and binarize the result with a $0.5$ threshold. 

\paragraph{Individual vertebra segmentation.}
The goal of this step is to obtain an individual vertebra segmentation from a detected vertebra location. 
We use as input a crop of the CT image centered at the given 3D vertebra location with an auxiliary channel in which the location is encoded \citep{payer2020coarse}. 
This channel contains the 3D location converted into a 3D probability map using a Gaussian kernel with $\sigma = 20$. The output is a one-channel probability map which is then binarized with a $0.5$ threshold to obtain the final binary mask. 
The input patch size is $128 \times 128 \times 128$ which allows to cover any full shaped vertebra in an arbitrary orientation. The $\lambda$ in Eq.~\ref{loss:dice_l2} is set to $20$ during training. 


\subsection*{B. Vertebrae statistics}

\renewcommand{\thetable}{B\arabic{table}}
\setcounter{table}{0}

\paragraph{Vertebrae volume regressor}
Vertebrae volumes vary across the cervical, thoracic and lumbar while the consecutive vertebrae volumes are heavily related. We learn the statistics of the vertebrae volumes to predict one from the neighboring (previous or next) one. To be specific, for each anatomic group (cervical, thoracic and lumbar), two linear regressors are trained to predict the vertebra volume given the previous and the next vertebra respectively:
$S_i = a * S_{i-1} + c_1$ and $S_i = b * S_{i+1} + c_2$.
We use the identification of the neighboring vertebra to select the specialized regressor. 
The learned coefficients are 
$(a= 1.03,  c_1= 1471 , b= 0.92, c_2= 497)$ for cervical,
$(a= 1.03,  c_1= 1354, b= 0.94, c_2= -140)$ for thoracic
and
$(a= 1.05, c_1= 981, b= 0.94, c_2= -269)$ for lumbar.

\paragraph{Inter-vertebral distance regressor}
The vertebrae locations are consistent and well structured, we learn two forms of the inter-vertebral distance statistics. One is the Gaussian distribution for each anatomic group, the other is linear regressors predicting the inter-vertebral distance given either the previous, next or both-side distances. 

For the learned coefficients of the Gaussian distributed inter-vertebral distance, we have
$(\mu = 16.77, \sigma = 2.18)$ for cervical,
$(\mu = 23.32, \sigma = 3.55)$ for thoracic,
and
$(\mu = 32.68, \sigma = 2.84)$ for lumbar.
The distance $G_i$ is considered as the distance between two consecutive vertebrae when
$\mu-3\sigma < G_i < \mu+3\sigma$.

To predict the inter-vertebral distance given the neighboring distances, we train three types of the regressors: using the both-side neighboring distances $G_i = m_1 * G_{i-1} + n_1 * G_{i+1} + K_1$, using the previous distance $G_i = m_2 * G_{i-1} + K_2$ or using the next distance $G_i = n_2 * G_{i+1} + k_3$. The latter two are intended for the top and bottom vertebrae who do not have the distances from both sides.
The learned coefficient can be found in Table.\ref{tb:gap_regressor}.

\begin{table}[h]
\centering
\scalebox{0.85}{
\begin{tabular}{|l|c|c|c|c|c|c|c|}
\hline
         & $m_1$ & $n_1$ & $k_1$ & $m_2$ & $k_2$ & $n_2$ & $k_3$ \\ \hline
Cervical & 0.55 & 0.45 & -0.08 & 0.92 & 2.40 & 0.98 & -0.13 \\ \hline
Thoracic & 0.57 & 0.44 & -0.24 & 0.93 & 2.29 & 0.97 & -0.07 \\ \hline
Lumbar   & 0.56 & 0.46 & -0.73 & 0.95 & 1.96 & 0.96 & 0.23  \\ \hline
\end{tabular}
}
\caption{Learned coefficients of the inter-vertebral distance regressors.}
\label{tb:gap_regressor}
\end{table}

We then compute the Mean Relative Error (MRE) between the relative distance from the locations and the distance from the regressor. The distance is considered as normal when $\mu_{mre} - 3*\sigma_{mre} < MRE < \mu_{mre} + 3*\sigma_{mre}$. The learned $\mu_{mre}$ and $\sigma_{mre}$ for each anatomic group can be found in Table. \ref{tb:mre_coeff}.

\begin{table}[h]
\centering
\scalebox{0.85}{
\begin{tabular}{|l|c|c|c|c|c|c|}
\hline
\multirow{2}{*}{} & \multicolumn{2}{c|}{both sides} & \multicolumn{2}{c|}{previous} & \multicolumn{2}{c|}{next} \\ \cline{2-7} 
                  & $\mu_{mre}$  & $\sigma_{mre}$   & $\mu_{mre}$  & $\sigma_{mre}$ & $\mu_{mre}$   & $\sigma_{mre}$   \\ \hline
Cervical          & 9.13           & 2.86           & 10.20          & 2.05         & 12.13        & 3.36       \\ \hline
Thoracic          & 2.42           & 1.43           & 3.96           & 1.56         & 4.17         & 0.88       \\ \hline
Lumbar            & 2.04           & 0.93           & 4.45           & 1.55         & 5.16         & 1.71       \\ \hline
\end{tabular}
}
\caption{MRE mean and standard deviation of the inter-vertebral distance regressors.}
\label{tb:mre_coeff}
\end{table}

\end{sloppypar}
\end{document}